\documentclass{sigchi}

\CopyrightYear{2018}
\setcopyright{acmlicensed}
\doi{https://doi.org/10.1145/3173574.3173951}
\isbn{978-1-4503-5620-6/18/04}\acmPrice{$15.00}
\conferenceinfo{CHI'18,}{April 21--26, 2018, Montr\'{e}al, Canada}
\acmPrice{\$15.00}
\usepackage{tikz}

\usepackage{balance}       
\usepackage[subtle]{savetrees}
\usepackage{graphics}
\usepackage{graphicx}
\usepackage{xcolor}
\usepackage[framemethod=TikZ]{mdframed}

\mdfdefinestyle{mystyle}{
    backgroundcolor=black!5,
    roundcorner=2pt
}
\usepackage[T1]{fontenc}   
\usepackage{txfonts}
\usepackage{mathptmx}
\usepackage[pdflang={en-US},pdftex]{hyperref}
\usepackage{color}
\usepackage{booktabs}
\usepackage{textcomp}

\usepackage{microtype}        
\usepackage{ccicons}          

\usepackage{todonotes}

\def\plaintitle{`It's Reducing a Human Being to a Percentage'; Perceptions of Justice in Algorithmic Decisions}

\def\emptyauthor{}
\def\plainkeywords{Authors' choice; of terms; separated; by
  semicolons; include commas, within terms only; required.}

\makeatletter
\def\url@leostyle{%
  \@ifundefined{selectfont}{
    \def\UrlFont{\sf}
  }{
    \def\UrlFont{\small\bf\ttfamily}
  }}
\makeatother
\def\pprw{8.5in}
\def\pprh{11in}

\definecolor{linkColor}{RGB}{6,125,233}
\hypersetup{%
  pdftitle={\plaintitle},
  pdfauthor={\emptyauthor},
  pdfkeywords={\plainkeywords},
  pdfdisplaydoctitle=true, 
  bookmarksnumbered,
  pdfstartview={FitH},
  colorlinks,
  citecolor=black,
  filecolor=black,
  linkcolor=black,
  urlcolor=linkColor,
  breaklinks=true,
  hypertexnames=false
}

\begin{document}

\title{\plaintitle}

\author{%
  \alignauthor{Reuben Binns\textsuperscript{1}, Max Van Kleek\textsuperscript{1}, Michael Veale\textsuperscript{2}, Ulrik Lyngs\textsuperscript{1}, Jun Zhao\textsuperscript{1} and Nigel Shadbolt\textsuperscript{1}}\\
      \affaddr{\textsuperscript{1}Dept. of Computer Science, University of Oxford, United Kingdom}\\
      \affaddr{\textsuperscript{2}Dept. of Science, Technology, Engineering \& Public Policy, University College London, United Kingdom}\\
      \email{\{reuben.binns, emax, ulrik.lyngs, jun.zhao, nigel.shadbolt\}@cs.ox.ac.uk, \{m.veale\}@ucl.ac.uk}
}

\maketitle


\begin{abstract}
Data-driven decision-making consequential to individuals raises important questions of accountability and justice. Indeed, European law provides individuals limited rights to `meaningful information about the logic' behind significant, autonomous decisions such as loan approvals, insurance quotes, and CV filtering. We undertake three experimental studies examining people's perceptions of justice in algorithmic decision-making under different scenarios and explanation styles. Dimensions of justice previously observed in response to human decision-making appear similarly engaged in response to algorithmic decisions. Qualitative analysis identified several concerns and heuristics involved in justice perceptions including arbitrariness, generalisation, and (in)dignity. Quantitative analysis indicates that explanation styles primarily matter to justice perceptions only when subjects are exposed to multiple different styles---under repeated exposure of one style, scenario effects obscure any explanation effects. Our results suggests there may be no `best' approach to explaining algorithmic decisions, and that reflection on their automated nature both implicates and mitigates justice dimensions.
\end{abstract}

\category{H.5.m.}{Information Interfaces and Presentation
  (e.g. HCI)}{Miscellaneous}{}{}


\section{Introduction}
Important decisions---about the allocation of jobs, loans, or insurance---are increasingly based on the predictions and classifications of computer models. By training such models on large datasets of previous customers or employees, organisations seek to more accurately estimate the risks and rewards of different actions and act accordingly---sometimes even automatically. Such practices are commonly referred to as `algorithmic decision-making', and have caused both celebration and consternation~\cite{mittech2017,john2014big,oneill2016}.

However, such decisions have significant consequences for those individuals directly affected by them~\cite{hildebrandt2008profiling}. While human decision-makers may exhibit errors of judgement and biases, we can at least demand that they rationalise their decisions and thereby hold them accountable. A significant body of research in legal and organisational psychology suggests that people do not only care about whether the outcome of a decision benefits them, but also whether it meets standards of justice~\cite{thibaut1973procedural, lind1988social,colquitt2015measuring}. In particular, information about decision-making processes plays an important role in justifying the decisions that are made. In their basic form, algorithmic decision-making systems do not naturally provide information relevant to justice judgements, and therefore, threaten the capacity for accountability inherent in human decision-making~\cite{burrell2016machine}. In relying on such systems, we therefore run the risk of important decisions being made unaccountably.

The challenge to justice has long been recognised in contexts where data-driven decision-making is used in government administration~\cite{schwartz1991data}. The increasing use of predictive systems by both private and public bodies has prompted regulation aimed at rendering them more transparent. The European Union's General Data Protection Regulation (GDPR) (and similarly, the preceding 1995 Directive) requires organisations deploying certain systems to provide affected individuals with `meaningful information about the logic' behind their outputs~\cite{wachter2017right}. Similarly, the United States' Equal Credit Opportunity Act (ECOA) requires lenders to provide `statements of reasons' when applications are denied loans~\cite{hsia1978credit}.

These legal requirements give rise to HCI challenges concerning how the outputs of such systems should be communicated to affected individuals and other stakeholders. These challenges are not entirely new. The ability to provide information about how a system derives its predictions or classifications has long been recognised in the AI, expert systems, and HCI communities~\cite{clancey1983epistemology,neches1985enhanced,dourish1997accounting}. However, recent work has addressed the novel challenges of complex machine learning models, whose logic and outputs are otherwise harder to explain. So-called `model-agnostic' approaches aim to explain the outputs of any classifier, regardless of the machine learning algorithm used to train it~\cite{ribeiro2016should}.

The potential for, as well as the limitations of, such approaches to address these regulatory requirements have been noted~\cite{edwardsveale, barocasselbst2017, weller2017challenges}. By implementing these explanation facilities within their algorithmic decision-making systems, organisations may be able to meet their obligation to provide `meaningful information' about how their decisions are made. However, evaluation of the human factors of such systems has primarily tested their utility in supporting \emph{expert} decision-making (for instance, in clinical diagnostic settings~\cite{bussone2015role}). The kind of meaningful information required by experts is likely to differ from that required by individuals affected by automated credit scores, insurance pricing or hiring decisions, for the purposes of accountability as envisaged by the relevant regulatory regimes.

This study aims to provide a preliminary exploration of the ways in which a range of novel explanation approaches might serve regulatory goals of rendering algorithmic decision-making more fair, accountable, and transparent. In particular, what effects do \emph{explanations} have on people's perceptions of algorithmic decisions with respect to these regulatory aims? Do different explanation approaches meet these aims in different ways? To explore these questions, we conducted a series of studies, both lab-based and online, examining people's responses to a series of scenarios involving algorithmic decisions accompanied by different explanation styles inspired by recent work on fairness, accountability and transparency in machine learning.

\section{Background}

The questions raised above are connected to multiple distinct areas of research, including: the psychology of justice perceptions regarding decision-making; the design of interpretable models in machine learning and HCI; and interdisciplinary work at the intersection of machine learning, law and social sciences.

\subsection{Interpreting intelligent systems} \label{interpretableML}
The ability to interpret a system's outputs has long been a core theme of research into intelligent systems. As a general principle, systems should be able account for their own operation in ways that help users understand how their tasks are being accomplished~\cite{dourish1997accounting,shneiderman1997direct}. Early knowledge representation and reasoning systems typically aimed to meet this requirement by  producing explanations for their outputs, for instance by presenting derivations from rules in rule-based architecture~\cite{teach1981analysis, clancey1983epistemology}, or graphical representations of nodes and edges in the case of Bayesian networks~\cite{lacave2002review}. Explanation facilities could be further augmented by explicitly modeling the knowledge engineering and design process and revealing information about them to end-users~\cite{neches1985enhanced}. Inspired by different theoretical models of argumentation, various explanation types have been explored and tested, with the style of explanation provided affecting the extent to which end-users understand the decisions of expert systems~\cite{moore1991reactive,lim2009and}.

Various research suggests that users---again, primarily experts---may in certain cases systematically over-rely on the outputs of intelligent systems~\cite{goddard2011automation, Dzindolet:2003bl}, while in others they may systematically distrust systems, despite reassurances as to their accuracy~\cite{onkal2009relative,fildes2009effective}. However, the provision of explanations can affect levels of trust and acceptance of algorithmic decisions. In some cases, explanation increases trust and reliance~\cite{johnson1993explanation,ye1995impact,antifakos2005towards,herlocker2000explaining,pu2006trust}, but in others an explanation may have the opposite effect if the level of detail it contains is deemed insufficient~\cite{bussone2015role}.

The increasing popularity of machine learning models as decision making tools presents novel challenges for explanation, as it may not be as straightforward to extract the embedded `knowledge' behind their outputs~\cite{tickle1998truth}. This problem is particularly pronounced for complex models, like many-layered neural networks, but even simple decision tree models can quickly become uninterpretable if they have too many branches. As users form mental models~\cite{norman1983some} of machine learning systems over time, such long-term exposure may lead to greater understanding~\cite{tullio2007works} and greater trust and reliance~\cite{muir1994trust}. But in the absence of transparency from platform providers, end-users may develop folk theories of algorithms, leading to a diverse range of strategies and countermeasures with varying degrees of effectiveness~\cite{eslami2016first,devito2017algorithms}.

Recent work on interpretable machine learning has focused on providing primarily \emph{pedagogical}~\cite{tickle1998truth}, \emph{local} explanations for the outputs of ML models: pedagogical in the sense that the explanations teach something about how the model works rather than attempting to represent it directly, and `local' in that they focus on how a specific output was derived. Approaches such as Local Interpretable Model-Agnostic Explanations (LIME)~\cite{ribeiro2016should} and Quantitative Input Influence (QII)~\cite{datta2016algorithmic} enable an end-user to see a list of features which contributed to the output, along with the strength and direction of the contribution to the outcome class.\footnote{Layer-wise Relevance Propagation/Deep Taylor Expansion~\cite{montavon2017explaining}, decomposes the innards of deep learning systems to provide similar results to LIME (without directionality), and is also relevant here.} Case-based explanations provide examples from the model's training data which most closely resemble the output in question~\cite{cunningham2003evaluation,nugent2005case}, while `demographic' explanations provide statistics on the outcome classes for relevant demographic variables~\cite{ardissono2003intrigue}.

While these different explanation \emph{styles} suggest promising ways to explain the outputs of any classifier, explanation quality---or `interpretability'---does not have a formal definition or a standard evaluation methodology shared amongst machine learning researchers~\cite{doshi2017towards,lipton2016mythos}. Explanation serves different functions in different contexts, and its evaluation is therefore context- and purpose-dependent. To this end, various human subjects experiments have been undertaken, exploring whether particular explanations result in better performance on a given end-task, in comparison to a no-explanation baseline. Contexts studied include aiding medical experts in diagnosis, increasing test performance of users of personalised learning environments, or enabling network security analysts to correctly identify incidents (e.g.~\cite{kim2013inferring,williams2016axis,ehrlich2011taking}).

The use of various explanation styles has been been widely explored in the field of recommender systems. Tintarev and Mastho identify seven purposes for recommender system explanations, namely: transparency, scrutability, trust, effectiveness, persuasiveness, efficiency and satisfaction~\cite{tintarev2015explaining}. Explanations can also be used to \emph{justify} or \emph{describe}~\cite{vig2009tagsplanations}. While early recommender system explanation approaches provided a uniform explanation style for single-source collaborative filtering ~\cite{herlocker2000explaining}, more recent work explores how to derive explanations for recommender systems based on `hybrid' multiple sources ~\cite{kouki2017user} and matrix factorisation ~\cite{rastegarpanah2017exploring}. However, a reported need for explanation may not always correspond to differences in behaviour or performance; in a study of news recommender systems, end-users expressed a desire for explanations, but the number of news items they opened did not change when provided with reasons for their recommendations~\cite{ter2017news}.

The use of local, pedagogical explanation facilities has not yet been tested in the contexts affected by the aforementioned emerging regulatory requirements, such as the automated evaluation of people for loans, hiring, and insurance. While various tools designed to help decision-makers identify and correct discrimination in data mining exist, these are aimed at data scientists implementing systems rather providing information to the individual decision-subjects affected by them~\cite{berendt2014better,Pedreshi:2008ej,Hajian1:2012,feldman2015certifying,grgic2016case}. The potential for local, pedagogical explanation systems to provide justice-related information, fulfilling the policy goals of transparency, accountability and fairness, has recently been noted by computer scientists and law scholars~\cite{edwardsveale, barocasselbst2017, weller2017challenges}. It has been suggested that organisations might rely upon these explanation facilities to fulfill legal duties to provide meaningful information about the logic of specific system outputs to affected individuals.


\subsection{Perceptions of justice regarding decision-making}

In order to explore how such explanation facilities might serve the aims of justice in algorithmic decision-making, a clearer understanding of the psychological aspects involved in justice perceptions is required. To this end, it is worth considering the extant literature on perceptions of justice regarding \emph{human} decision-making. Decisions made about people with significant effects, such as court decisions, hiring, firing and promotion in a workplace, or the allocation of financial products, are often contentious and require a higher burden of accountability than other decisions; they should meet the standards of `bureaucratic justice'~\cite{mashaw1985bureaucratic}. Early work in this area emphasised that perceived levels of justice of a decision outcome are separate from purely self-serving rationalisations of a decision outcome; an individual might be negatively affected by a decision whilst still thinking it is just~\cite{thibaut1973procedural, lind1988social}. 

According to Colquitt and others, justice perceptions can be broken down into several aspects or antecedents~\cite{colquitt2001justice}. \emph{Procedural} justice concerns the processes, logic and deliberation behind a decision. \emph{Distributive} justice concerns the allocation of positive and negative outcomes in a decision context and whether they are distributed equitably or deservedly amongst the affected population given their circumstances, performance or contributions. \emph{Interactional} justice concerns the extent to which the affected individual is treated with dignity and respect by the decision-makers. Finally, \emph{informational} justice pertains to the information and explanations provided for decisions; are they candid, thorough, and tailored to individual needs? These aspects of justice are distinct but correlated~\cite{colquitt2001justice}. Receiving a thorough explanation (informational justice) is important in helping people to assess whether the decision-making procedure is just (procedural justice)~\cite{colquitt2001justice}. In turn, decisions perceived to be procedurally just are more likely to be perceived as distributively just~\cite{van2014role}. Such findings provide a rationale for giving individuals rights to information about significant decisions; requiring decision-makers to explain their decisions should promote informational justice, which should help people assess procedural justice and distributive justice in turn.
%
%

This rationale might plausibly apply to the regulatory regime around algorithmic decision-making. Requiring organisations to explain the logic behind their algorithmic decision-making systems (informational justice) enables affected individuals to assess whether the logic of the system is just (procedural justice), which in turn might moderate their assessments of fairness of the decision outcomes (distributive justice). In so far as these notions of justice capture the aims of the regulatory requirements, they may provide appropriate ways of measuring the adequacy of different explanation systems for these purposes.


\subsection{Key questions and contributions}

This work builds upon and contributes to these disparate areas of research. Work on perceptions of justice reveals much about the role of explanations in ethical assessments of human decisions, but the applicability of these findings to algorithmic decisions and the range of proposed explanation systems proposed for them remains largely unexplored. A notable exception is Kizilcec, who studied the effect of levels of transparency on perceptions of procedural justice~\cite{kizilcec2016much}. However the nuances of the different explanation approaches recently developed in interpretable ML have not yet been explored in relation to perceptions of justice. We also do not yet know much the extent to which findings on correlations between perceptions of justice from human decision-making contexts hold in the context of algorithmic decisions.

Furthermore, as research into practically achievable explanations for machine learning systems accelerates, researchers are turning to consider which explanation styles are more desirable, rather than simply possible to implement~\cite{lipton2016mythos,doshi2017towards}. This itself is a loaded question: an explanation style that artificially inflates certain notions of justice by playing on cognitive biases might suffer from the same anti-paternalist critiques levied against proponents of `nudge' philosophy~\cite{mitchell2004libertarian}. Understanding how explanation styles might influence and interact with justice perceptions is therefore an important preliminary to further development of ML explanation tools.

With these issues in mind, our key questions are:

\begin{enumerate}
  \item How do explanations for algorithmic decisions affect justice perceptions regarding algorithmic decisions? In particular, do the positive correlations observed between informational, procedural and distributive justice in human decision-making settings also hold in algorithmic decision-making settings?
  \item How do \emph{different styles} of explanation affect such justice perceptions?
\end{enumerate}

\section{Study Design and Methodology}

In investigating these questions, we combine several methodological aspects of the related work from justice perceptions and intelligent systems research. Many of the previously mentioned machine learning model explanation systems have not been subjected to user evaluations, and those which have often only test people's ability to predict a model's outputs, or answer questions about its logic (e.g. ~\cite{lim2009and}). For present purposes, we are not concerned about how faithfully an explanation facility imparts actual model logic to a user.\footnote{Although we acknowledge this is an important challenge if explanations are to be more than just comforting stories, as cautioned by Lipton~\cite{lipton2016mythos}.} Rather, we are interested in how the provision of various kinds of information about a model's outputs might affect justice perceptions as measured in the psychological literature. Much of this research is based on field surveys of individuals with personal experience of being affected by decisions (e.g.~\cite{van2014role}), although controlled experiments involving simulated decision-making and scenario-based methods have also been used~\cite{van1997judge}. 

Given our focus on the effects of information provision and explanation styles on perceptions of fairness, we conducted a set of experimental studies to elicit people's responses to a range of plausible algorithmic decision scenarios and associated explanations. While this may lack the ecological validity of field surveys, it gives us the ability to engage people with a wide range of novel explanation styles covered in \ref{interpretableML}), and have yet to be implemented alongside real machine learning systems currently deployed `in the wild' to any significant degree. By testing out proposed explanation styles with fictional scenarios, we hope that designers can gain insight into user concerns, and perhaps create more meaningful explanation interfaces to meet the upcoming legal requirements.

\subsection{Scenarios and application contexts}

We selected 5 application contexts in which to situate the fictional scenarios. The main criteria for context selection were: a relatively common interaction for the target population (a large proportion has experienced or will experience it at some point in their life); likely to involve machine learning models for algorithmic decision-making (at present or in the near future); have significant economic or practical effects on the decision-subject; and would possibly fall under the aforementioned GDPR regulation on automated decisions.\footnote{The hurdle that a decision must be `based \emph{solely} on automated processing' is here read less restrictively than legal analysis has indicated~\cite{wachter2017right,edwardsveale}.} After consulting relevant literature on how ML systems are being applied in each context (e.g.~\cite{john2014big,lessmann2015benchmarking,john2014big,o2017weapons,thomas2017credit}), we settled on the following contexts:

\begin{enumerate}
\item Applying for a personal financial loan;
\item Applying for a promotion at work;
\item Car insurance premiums dynamically priced based on personal details and driving behaviour measured by a telematic sensor (as described in e.g.~\cite{vaia2012vehicle});
\item Passengers on over-booked airline flights being selected for re-routing;
\item Freezing of bank account due to activity suspected as associated with money laundering;
\end{enumerate}

For each context, we created fictional cases in which an individual has had a decision made about them automatically. These were inspired by now-classic algorithmic `war stories'~\cite{edwardsveale} (see, e.g.~\cite{oneill2016,pasquale2015}). Scenarios and cases were adapted where necessary to fit the context from which the in-person participants would be recruited (a city in the United Kingdom). Because explanations are most likely to be requested in response to a decision which has a negative outcome for the individual, and considerations of procedural justice have been found to be more strongly felt in such cases~\cite{brockner1996integrative}, each scenario ended in a negative outcome for the individual.
 
\subsection{Explanation styles}

In order to derive a set of explanation styles to test, we reviewed a range of research, including both technical work on interpretable machine learning models, and legal discussions on the requirements of existing and forthcoming regulations on transparency of algorithmic decisions (e.g.~\cite{wachter2017right,barocasselbst2017,edwardsveale}. Our aim was to find explanation styles which could plausibly meet or exceed the regulatory requirements regarding transparency of automated decisions, in particular the requirement that organisations provide `meaningful information about the logic involved' in an automated decision (GDPR art 15(h)). We aimed to condense the many different proposals into a smaller set of explanation styles based on the kinds of information they would likely present to the end-user. We included only explanation methods which were model-agnostic; i.e. are applicable to any kind of learning algorithm, such as LIME~\cite{ribeiro2016should}, rather than methods which only work for particular ML models (e.g. those restricted to random forests~\cite{strobl2007bias} or neural networks~\cite{montavon2017explaining}). This review resulted in 4 categories of promising explanation style:

\begin{enumerate}
\item \textbf{Input Influence}: Presents a list of input variables alongside a quantitative measure of their `influence', positive or negative, on a decision (e.g.~\cite{datta2016algorithmic,zeng2017interpretable})
\item \textbf{Sensitivity}: For each input variable used in a decision, sensitivity analysis shows how much the value of that variable would have to differ in order to change the output class (not to be confused with the notion of sensitivity used in ML evaluation) (e.g.~\cite{samek2016evaluating,barocasselbst2017}).
\item \textbf{Case-based}: Presents a case from the model's training data which is most similar to the decision being explained (e.g.~\cite{donal2003})
\item \textbf{Demographic}: Presents aggregate statistics on the outcome classes for people in the same demographic categories as the decision-subject, such as age, gender, income level or occupation (e.g.~\cite{ardissono2003intrigue})
\end{enumerate}

\begin{figure*}[ht!]
\includegraphics[width=\linewidth]{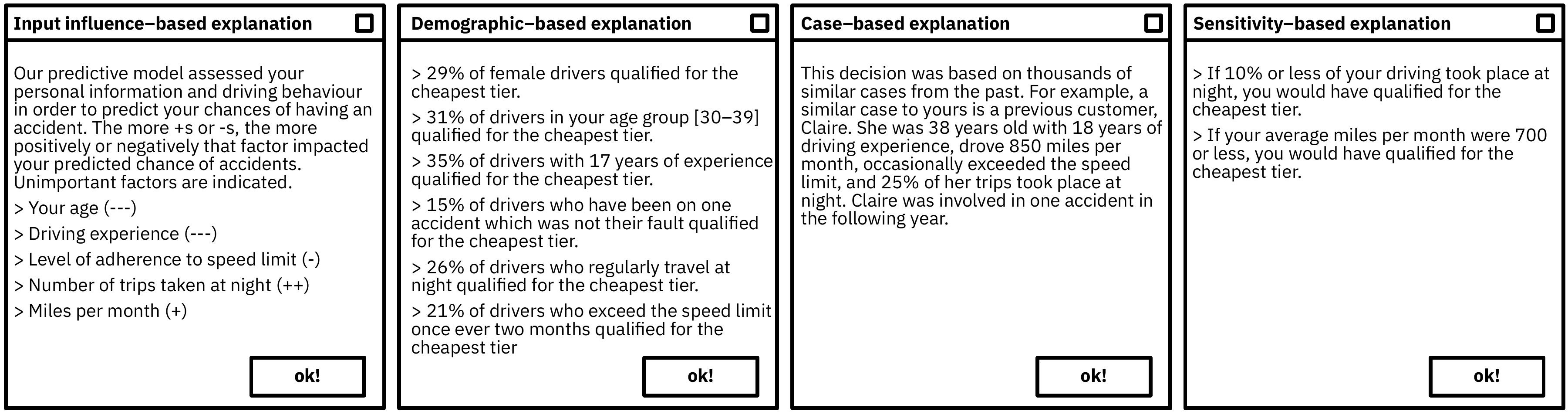}
\caption{Illustrative examples of the different explanation types presented to participants: input influence, sensitivity, demographic and case based.}
\label{fig:explns}
\end{figure*}

These explanation styles were tested and refined in a small-scale informal design phase. Each explanation system imparts different information, which is presented in various ways in the literature, including text~\cite{donal2003}, graphs~\cite{ribeiro2016should} and bullet points~\cite{zeng2017interpretable}. In order to control for these differences in representation, we chose to use purely textual explanations. We informally tested each explanation style by varying the wording and information dimensions. We found that people generally did not pay attention to explanations independent of scenario descriptions. Through iterative testing we found that the use of dialog boxes focused attention on explanations (see Figure \ref{fig:explns}). These were deployed and investigated in depth in an in-person lab study and two online experiments. All phases of each study were approved by relevant University ethical review procedures.

\begin{figure}[htbp]
\begin{mdframed}[style=mystyle]
  \begin{scriptsize}
    \textbf{Car insurance scenario:}
A car insurance company provides customers with personalised prices based on their attributes and driving behaviours, measured through a telematic sensor installed in the car. Their system for setting prices is based on a computer model, which predicts how likely an applicant is to have an accident and make a claim. The computer makes its predictions based on data the system has collected about thousands of other drivers.

Based on its ongoing analysis of the driver's chances of having an accident, the system will automatically set their monthly premium. The cheapest premium tier, given to those drivers who are judged by the system as the safest, is 20 per month.

\vspace{2mm}
\vfill

\textbf{Sarah}

Sarah is a customer of the insurance company. She is:
\begin{itemize}
\setlength\itemsep{-0.5em}
\item 35 years old
\item Has been driving for 17 years
\item Been in an accident once which was not her fault
\item Drives 800 miles a month on average
\item Exceeds the speed limit on average once every two months
\item 20\% of Sarah's driving takes place at night
\end{itemize}
\vspace{2mm}
\vfill

Based on this information, the computer system has not qualified Sarah for the cheapest tier of insurance premium. The insurer provides Sarah with the following information about the computer's decision.
\vspace{2mm}
\vfill
"This decision was based on thousands of similar cases from the past. For example, a similar case to yours is a previous customer, Claire. She was 38 years old, with 18 years of driving experience, drove 850 miles per month, occasionally exceeded the speed limit, and 25\% of her trips took place at night. Claire was involved in one accident in the following year."
\vspace{2mm}
\vfill
    \textbf{Please rate your agreement with the following statements}
\begin{enumerate}
\setlength\itemsep{-0.5em}
\item \emph{Agreement}: `I agree with the decision'
\item \emph{Understanding}: `I understand the process by which the decision was made'
\item \emph{Appropriateness of factors}: `The factors considered in the decision were appropriate'
\item \emph{Fair process}: `The decision-making process was fair'
\item \emph{Deserved outcome}: `The individual deserved this outcome given their circumstances or behaviour'
\end{enumerate}
  \end{scriptsize}
  \end{mdframed}
  \caption{Example of a scenario (car insurance), case description (Sarah), explanation (Case-based), and questions}
  \label{ataglance}
\end{figure}

\subsection{Justice constructs}\label{justiceconstructs}
The justice constructs used in all three studies were based on those developed in the psychology of justice research, summarised by Colquitt et al~\cite{colquitt2015measuring}. Participants were asked to rate their agreement with five statements on a 5-point Likert scale (see statements at the bottom of Figure \ref{ataglance}). Statements were drawn and adapted from previous studies of human decision-making to fit scenarios in which the decision was being made by an automated algorithmic system. Preliminary testing found that certain questions from Colquitt's scale, particularly those relating to \emph{interactional} justice such as `Have you been able to express your views and feelings during those procedures?' were difficult to interpret. This was partly due to the hypothetical nature of the scenarios, and partly due to being too human-specific to apply to a computer-based interaction. The remaining questions pertained to informational, procedural and distributive justice. Statement 2 (\emph{understanding}) aimed to capture informational justice~\cite{greenberg1993social}; statements 3 and 4 (\emph{appropriateness of factors, fair process}) pertained to procedural justice; and statement 5 (\emph{deserved outcome}) addressed distributive dimensions~\cite{colquitt2015measuring}. We also found that people tended to interpret these justice construct statements as a proxy for the simpler question of whether or not they \emph{agreed} with the decision; to avoid this, an additional statement about agreement (1, \emph{agreement}) was also included prior to the others.


\subsection{Phase 1: Lab study}
The first study consisted of in-person semi-structured interviews focused around a series of fictionalised cases as described above. The purpose of this study was to gain in-depth insights into how people interpret, evaluate and reason about algorithmic decision-making in the range of contexts and explanation styles mentioned above.

19 participants were recruited through paper flyers placed in various permitted locations in a small city in the UK (including caf\'{e}s and restaurants, churches, libraries, shops, and museums), and through official social media accounts of the research lab and personal accounts of the researchers. Participants were then invited to the lab where the study occurred, which was set up with a standard PC keyboard and mouse, and a 27-inch display with a screen resolution of 1920x1200.

Participants were presented with 3 cases for each of the 5 contexts, (15 cases in total) via a web-based interface. Cases were presented with a brief description of the individual, the decision outcome, the explanation (except in the control condition) and an illustrative image depicting the context. The cases were used to prompt reflection on various aspects of the decision, and any explanatory information provided, using a concurrent think-aloud process~\cite{nielsen2002getting}. Participants were asked to consider their agreement with the five measures described in section~\ref{justiceconstructs} and Figure~\ref{ataglance}. Agreement was indicated on a 5-point Likert scale, ranging from `strongly agree' to `strongly disagree'. Participants were asked to verbalise their responses to each case and explain their thought processes for each response. Each case was presented with one of the 4 explanation styles described above, or a control condition in which no explanation was provided. Participants were assigned pseudonymous identifiers to be associated with data generated during the experiment.


Audio recordings taken from the interviews and think-aloud processes were transcribed, segmented and labelled by each case and explanation style, and any personally identifying information was redacted. Thematic analysis was undertaken to identify common themes in participants' interpretation, evaluation and reasoning in response to cases. Thematic codes were independently developed by three coders, after which a single set was jointly agreed upon. The data were then re-coded using the new set of codes. 

\subsection{Phase 2: Online studies}
Following on from the lab study, two online studies were conducted in order to generate quantitative data to test the following hypotheses:

\begin{enumerate}
\item Do different explanation styles result in differences in perceived levels of justice?
\item Do the positive correlations observed between informational, procedural and distributive justice in human decision-making settings also hold in algorithmic decision-making settings?
\end{enumerate}

Both online studies consisted of pared-down versions of the protocols used in phase 1. Participants were recruited via the study platform Prolific Academic,\footnote{\url{www.prolific.ac}} and filtered to include only individuals over 18 based in the UK, in order to maintain similarity with the participants recruited in phase 1.

A between-subjects design was used in order to test differences between responses under different explanation conditions. On the basis of described responses to the various explanation styles in the lab study, it was hypothesised that the ability to make direct comparisons between explanation styles might affect justice perceptions. We therefore also devised a follow-up within-subjects experiment, to test whether exposure to multiple explanation styles for a single case would have different effects than repeated exposure to a single explanation style.

\subsubsection{Between-subjects study}
325 participants were randomly assigned to 1 of 5 conditions (n = 65 in each group). Each group was presented with a selection of 12 cases used in experiment 1, accompanied by one of the 4 different explanation styles or a control condition featuring no explanation.\footnote{The airline context was excluded at this stage, because experiment 1 had revealed that participants almost universally judged the cases to be very unfair and undeserved, thus preventing any useful conclusions to be drawn from this context.} Participants were asked to indicate their agreement with the 5 measures described in phase 1 using the same Likert scale. 

\subsubsection{Within-subjects study}
The within-subjects study involved 65 participants. Only the loan and insurance cases were included in this design, as the other scenarios could not be adapted to fit into a realistic hypothetical scenario in which four different explanations would be presented for a single case. Participants were therefore presented only with cases drawn from the Loan and Insurance contexts, in which the individual has received negative decision outcomes from 4 different lenders/insurers. Each decision was explained using a different style (in a random order), allowing participants to directly compare multiple styles applied to an individual subject.

Data from both studies were analysed with Spearman's rho correlations to understand if justice constructs correlate in the same way as reported in the aforementioned psychology of justice literature, and an ANOVA  with Tukey's post-hoc paired tests to understand how explanation styles relate to each other in light of different justice constructs. Details on both methods are presented further below alongside the analysis.

\section{Results}

\subsection{Participant information}








The lab study featured 19 participants, 11 male, 8 female, with an average age of 28.8 (range = 21-60, sd = 10.2). They had a range of educational attainment, comprising: high-school / A-levels (4), Bachelors (7), Masters (7) and PhD (1). In terms of familiarity with the decision-making contexts featured in the study, 6 had experienced applying for a loan, 6 had sought a car insurance deal, 5 had experience of promotion at work, 8 had had their bank accounts frozen, and 1 had been bumped off a flight. Average time to complete the study was 34 minutes (sd = 6.56).

For the between-subject online study, 64\% were female, with an average age of 37.6 (r = 17-69, sd = 11.4); 48\% were in full-time employment, 26\% were unemployed, 21\% were in part-time employment (3\% listed 'other'). Education levels comprised high school (16\%), A-level (27\%), undergraduate degree (34\%), Masters (16\%), and Doctorate (0.4\%). 25\% had experienced applying for a loan, 34\% had sought a car insurance deal, 26\% had experience of promotion at work, 14\% had had their bank accounts frozen, and 2\% had been bumped off a flight. Average time to complete the study was 7.5 minutes (sd = 3.26).

For the within-subject online study, 64\% were female, with an average age of 39.77 (range = 24-70, sd = 11.8); 66\% were in full-time employment, 14\% were unemployed, 19\% were in part-time employment (1\% listed 'other'). Education levels comprised high school (20\%), A-level (18\%), undergraduate degree (42\%), Masters (17\%), and Doctorate (3\%). 26\% had experienced applying for a loan, 30\% had sought a car insurance deal, 32\% had experience of promotion at work, and 9\% had had their bank accounts frozen. Average time to complete the study was 8.1 minutes (sd = 3.59).

\subsection{Qualitative Results: Themes and reflections}

The think-aloud responses from the lab study resulted in five major themes: the lack of human touch; interpretations of the system's reasoning; the use of statistical inference; the degree of actionability in an explanation; important aspects which were seen to be unaccounted for by the system; and the meaning and relevance of moral concepts. Many of these themes occurred across different explanation styles and scenario types; although some reflections were particularly concentrated around certain explanation styles and scenarios, which are noted below.

\subsubsection{The (lack of) human touch}
An initial reaction from many participants concerned the indignity or `weirdness' of algorithmic decision-making compared to human decision-making. A series of comments reflected upon the ways in which the very use of an algorithmic decision-making system could be impersonal and dehumanising for the subject of the decision. In response to a case in which an employee had been automatically rejected for a promotion, participant \emph{DC} remarked:

\begin{quote}
`[sarcastically] This really makes you feel like a valued employee! When a machine or some kind of system makes a decision about someone based on some kind of data points, then it's impersonal' --DC
\end{quote}

Others considered it problematic that with an automated system, `there's no sense of negotiation' (\emph{DR}), and no opportunity for `human interaction' (\emph{CS}). In contrast to this apparent lack of humanity (or perhaps to counteract it), some participants attempted to humanise or anthropomorphise the system~\cite{nass1993anthropomorphism}, one participant for instance describing it as `rude' (\emph{OR}).

For the same participant, the specificity of the figures used in some explanations (particularly, \emph{input influence} and \emph{sensitivity}) lead her to conclude that the computer must be using arbitrary thresholds, which in itself seemed `mean':

\begin{quote}
`Oh that's so mean! [...] I can't do the maths, but why is it so specific? Hmmm. I don't understand. I don't know why the cut-off is like that.'
--OR
\end{quote}

Such remarks suggest that if algorithmic decisions are held to norms of social behaviour, \emph{interactional} justice may be a relevant dimension for evaluation.

\subsubsection{Interpreting the system's `reasoning'}

Many of the reactions concerned attempts to decipher the logic of the algorithm. Some subjects made their assessments on the basis of whether the system approximated their own reasoning or knowledge:
\begin{quote}
`The computer [...] used the same sort of reasoning that I did I guess' --MP
\end{quote}
\begin{quote}
`Yeah I understand this decision [denying a young male cheap car insurance], because young males are much more likely to have accidents' --PC
\end{quote}

Others tried to articulate the system's reasoning as a series of rules and consequences:
\begin{quote}
`Breaking the speed limit is breaking the law, therefore you are proving that you are not capable of meeting the standards for the cheapest insurance, which is presumably: ``thou shalt obey all the laws''.' --AS
\end{quote}

In some cases, while the individual premises seemed understandable, their relevance to the reasoning behind the overall decision was not:

\begin{quote}
`It seems like I can understand every one of them, but if put into the picture [...] each point is relevant, but I don't know [how] it puts into the big picture' --OR
\end{quote}

Faced with seemingly inexplicable reasoning processes, some participants concluded that the system must be simply `making it up' (\textit{OR}) and therefore on the face of it unfair.

\subsubsection{Acceptability of statistical inference}

In addition to concerns about indignity and opaque reasoning, participants commonly reflected on the (un)acceptability of generalisation and statistical inference as a basis for prediction and decision-making in general. Issues raised in this regard were the importance of scientific rigour, including the use of sufficient sample sizes:

\begin{quote}
`I don't know how many previous customers they're basing it on...' --VI
\end{quote}
\begin{quote}
`I'm gonna assume that it looked at more than just John and was like: `You're like these people and these people didn't perform very well so we're not gonna give you a promotion'' --MP
\end{quote}

However, many other participants expressed strong objections to the use of statistical inferences to make judgements about people in itself:

\begin{quote}
`This is just simply reducing a human being to a percentage. It's not taking any of his actual ability, success or whatever into account. It's just saying `tough luck', you can only expect to be successful 20 percent [of the time]' --UT
\end{quote}

\begin{quote}
`This is saying your gender determines how you act, how you operate, and that's based on sheer probability.' --BF
\end{quote}

\begin{quote}
`Perhaps it's unfair to make the decision by just comparing him to other people and then looking at the statistics, he isn't the same person.' [...] `They [...] seem like [...] just random stats, not reasons for why you'd make a decision' --MP
\end{quote}

Some emphasised that an individual who seems similar to past individuals may in fact have extenuating circumstances, particularly in response to \emph{case-based} explanations:

\begin{quote}
`Just because Joel performed badly in the end doesn't mean that Jing [will] perform badly, because there could've been other influencing factors' --MP
\end{quote}

In addition to comparisons between different individuals, participants also commented on the acceptability of judging a single individual's future behaviour on the basis of their past behaviour. However in this case, some participants were more comfortable with inductive inferences; \textit{AS} agreed with the computer's decision to deny a loan applicant with existing debts because they had already `proven not to be able to pay it back'.

\begin{figure*}[tb!]
\includegraphics[width = \linewidth]{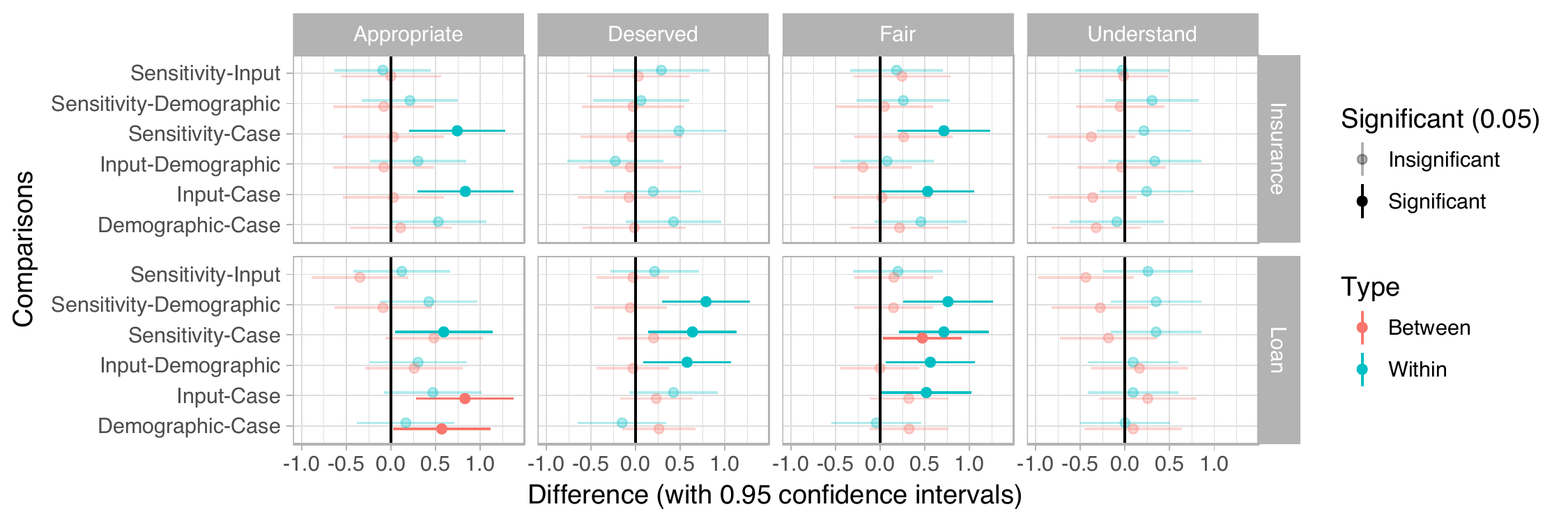}
\caption{Tukey's post-hoc paired tests showing comparative differences between answers in different scenarios by question (horizontal facet), scenario (vertical facet), study design (color) and explanation type (y-axis). Insignificant results are displayed in lower transparency for convenience.}
\label{tukey}
\end{figure*}

\subsubsection{Actionability}
In judging the appropriateness of certain features, or the fairness of a decision, issues of personal responsibility and the possibility of acting differently were frequently raised. Some participants were happy to endorse negative decisions if they were able to identify a reasonable alternative course of action available to the participant which would resulted in a different outcome, as in \emph{sensitivity}-based explanations:

\begin{quote}
`If he's not earning enough money for the company to feel they can give him a loan of 5,000 pounds, then that's fair. But they're also giving him an alternative in there; `if you had an extra 2,000 pounds income, you'd have been successful', or `if you'd asked to borrow [...] less, you'd have been successful'. Which is the right thing to do---reject him, but give him alternatives.' --MP
\end{quote}

Where an individual was deemed at fault for risky or otherwise undesirable past behaviour, participant \emph{AS} believed the decision was particularly justified:

\begin{quote}
`Drive more carefully [...] unless it is down to entirely unluckiness, you had a lot of accidents so you deserve it [being denied cheap insurance]' --AS
\end{quote}

Conversely, where a judgement was made on the basis of circumstances which were outside the individual's control, or the course of action suggested was deemed unrealistic, the outcome was seen as undeserved:

\begin{quote}
`I don't think she could have done anything [...] she didn't make any poor decisions that led to this, so I somewhat disagree that she deserved it' --PC
\end{quote}
\begin{quote}
`You don't choose where you're born, you don't choose your gender.' --BF
\end{quote}

Several responses suggested that explanations ought to help guide future action, something which was particularly lacking in the \emph{demographic} explanations:

\begin{quote}
`What do you do with `1.5\% of women have had their accounts frozen'? That gives you nothing, no information you can do anything with' --PC
\end{quote}

Although participant \emph{CS} was concerned that behaviour change in response to an algorithm could prove dangerous
\begin{quote}
`If you know it's on the back of an algorithm, it would incentivise people to work out how to game the algorithm, to find out what the algorithm is exactly doing' --CS
\end{quote}

\subsubsection{Unaccounted aspects}
Many criticisms of the decisions came down to things that were missing or unaccounted for. For some, it was the absence of important information:

\begin{quote}
`It should also base it on his credit score as well' --VD
\end{quote}
\begin{quote}
`I think the bank ought to know a bit more about you' --DR
\end{quote}
\begin{quote}
`We have no idea why John wasn't able to pay off the loan, he may have suddenly been made suddenly terminally ill' --UT
\end{quote}

In some cases, what was missing was not just information, but also complexity:

\begin{quote}
`If you just ask them more questions you could find out they're fine [...] or just make the algorithm more complicated' --DR
\end{quote}

In others, the missing element was a sense of proportionality. Responding to an \emph{input influence} explanation for denial of a loan, VI remarked:

\begin{quote}
` I don't know why [...] why he's got a minus 7 for existing debts. I think the information was relevant but I think it's kinda out of proportion' --VI
\end{quote}

\subsubsection{Meaning and relevance of moral concepts}

The meaning and relevance of moral concepts like `fairness' and `desert' in the context of a computer system was occasionally questioned by certain participants. Participant \emph{PC} struggled to evaluate the moral dimensions of computer-driven decisions:

\begin{quote}
`These last two questions [`Was the decision-making process fair?', `Did the individual deserve this outcome?'] are hard to answer... You can't really make a judgement about a computer, if you tell it to do something and it just does it.' --PC
\end{quote}

For \emph{DR}, here was a sense that the very notion of `fairness' may not be attributable to a system that is primarily designed to serve a different organisational goal like efficiency:

\begin{quote}
`Fair in this sense hasn't really come into it, this is like a cold decision [...] It's not fair at all really, but it's understandable from the perspective of the business' --DR
\end{quote}

One participant argued that a truly random system could also be seen as \emph{more} `fair', at least for certain definitions of fairness:

\begin{quote}
`I suppose the fair thing would be to just do it completely randomly [...] like a Harvey Dent definition of fair---from Batman, you know, Two Face?---"You know what's fair? Chaos is fair!"' --DR
\end{quote}

\subsection{Quantitative results}
To recap from the methodology, 325 participants from the \emph{Prolific Academic} platform were presented with different scenarios and different explanation styles, and asked to respond to questions that related to them on 5-point Likert scales. 


\subsubsection{Do justice correlations from human decision-making settings hold in relation to algorithmic decisions?}

According to the literature on justice perceptions above, we should expect to find correlations between \textit{fair process} and \textit{deserved outcome}, \textit{understanding} and \textit{fair process}, \textit{appropriateness of factors} and \textit{deserved outcome}, and \textit{appropriateness of factors} and \textit{fair process}~\cite{colquitt2001justice}. We were not expecting strong links between \textit{appropriateness of factors} and \textit{understanding} (as with a high level of understanding, one might find the factors inappropriate) or \textit{understanding} and \textit{deserved outcome} (as we suspect that judgments about deserved outcomes to be more dependent on inputs than on processes). To examine this, we calculated Spearman's rho correlation coefficients (with 0.05 confidence level bootstrapped intervals). The results (see Table \ref{table:constructs}) partially confirmed our hypotheses, as the stronger correlations were connected to the construct connections we believed would exist.
Surprisingly, we also found significant positive correlations between the two construct connections we were not expecting.

\subsubsection{Do different explanation styles result in differences in perceived levels of justice?} 

In order to examine contexts in which a decision subject is faced with a \emph{single} explanation style per decision, we conducted a between-subjects study. To examine contexts in which subjects are faced with \emph{multiple} explanation styles per decision, we conducted a within-subject study (like the qualitative study), which featured multiple explanation styles per decision.

We used ANOVA tests, followed by Tukey's post-hoc paired tests~\cite{tukey1949comparing}, to analyse the relative effects of different explanation styles within different study designs.\footnote{Some readers might have reservations over parametric tests being applied to ordinal data: we point to recent literature that has emphasised~\cite{norman2010likert} that Likert data are robust to parametric assumptions, and that `parametric methods can be utilized without concern for ``getting the wrong answer'''.} 

In the between-subjects study, explanation styles generally did not significantly affect justice perceptions, with the exception of fair process in loans only (loan F(3, 258) = 2.71, p = 0.046) and appropriate factors (F(3, 258) = 5.35, p = 0.001) showing significant between group differences.

In the within-subjects study---where multiple explanation styles were presented for the same decision---significant effects were observed across all questions and decision-making contexts. There was a significant effect of explanation styles on perceptions of fair process (loan scenario F(3, 260) = 7.52, p < .001; insurance scenario F(3, 260) = 4.5, p = .004), on the perception of appropriate use of factors (loan F(3, 260) = 3.312, p = .02; insurance F(3, 260) = 6.44, p = .0003), and on the perception of deserved outcome (only in the loans case, F(3, 260) = 7.31, p = .0001).

Tukey's post-hoc paired tests (see Figure \ref{tukey}) showed that case-based explanations result in lower perceptions of appropriateness, fair process perception, and (in the loans case) deservedness, consistently compared to sensitivity based styles and occasionally compared to other styles. This is an effect primarily observed, like most effects in the quantitative part of our study, in the \emph{within} subject study design, indicating that the act of comparison in a particular scenario is important for these differences to become apparent. Case-based explanations seem to have the most consistent negative impact on justice perceptions when presented alongside alternative explanation styles.

\begin{table}[ht!]
\centering
\small
\begin{tabular}{llrrr}
  \toprule
Q1 & Q2 & lower & rho & upper \\ 
  \midrule
  Fair & Deserved & 0.63 & 0.69 & 0.74 \\ 
  Fair & Appropriate & 0.53 & 0.60 & 0.66 \\ 
  Appropriate & Deserved & 0.43 & 0.51 & 0.58 \\ 
  Fair & Understand & 0.32 & 0.40 & 0.49 \\ 
  Appropriate & Understand & 0.31 & 0.40 & 0.48 \\ 
  Understand & Deserved & 0.28 & 0.36 & 0.44 \\ 
   \bottomrule
\end{tabular}
\caption{Spearman's rho correlations for justice constructs with bootstrapped (1000 rep) confidence interval at 0.02 level.}
\label{table:constructs}
\end{table}

\section{Discussion}
Faced with a range of algorithmic decision-making scenarios, people appear conflicted about how to respond to questions of justice. For some participants, the \emph{very idea} of an algorithmic system making an important decision on the basis of past data seemed unfair;

\begin{quote}
`She's been a victim of this computer system that has to generalise based on, like, somebody else ... what it should be looking for is ability to pay back the loan, not characteristics of others who couldn't pay back the loan' -WP
\end{quote}

Whereas for others, delegation of such responsibility to a computer, and the authority of statistical models, actually removed the relevance of questions of fairness altogether; the system `just does what it is supposed to', and as long as its inferences are accurate, it could be seen as `statistically fair' (\textit{PC}).

In this sense, while algorithmic decision-making implicates dimensions of justice, it may also mitigate them, in so far as people resist imputing morality to a computer system~\cite{friedman1992human, bryson2010robots}. On the other hand, people frequently interact with computers as if they were social actors~\cite{nass1994computers}; why stop when justice is involved? One participant described how she might have it both ways, simultaneously entertaining dual standards of justice, one juridical, the other computational;

\begin{quote}
`I would say I agree he deserved this outcome given the way this computer model has judged his behaviour, whether he actually deserved this outcome remains to be seen. If we were in a court of law, I would argue we don't know his circumstances, but given this computer model and the way it works it's deserved.' -WP
\end{quote}

Such attempts to delineate moral culpability between \emph{systems} and \emph{people}, or between different standards of fairness or evidence, may reflect a more general psychological need to see the world as morally coherent~\cite{clark2015moral}. The introduction of algorithmic systems seems likely to shake up previously stable categories and concepts, requiring adjustment and incorporation into one's moral world-view.

In terms of the value of providing explanations, and the relative importance of different explanation styles, the picture is complicated. On the one hand, people strongly engaged with the details of each explanation when discussing each case. This was borne out in the within-subjects experiment, where significant differences were observed in justice perceptions between different explanation styles---in particular that case-based explanation styles impacted negatively on justice-related judgements especially compared to sensitivity-based explanation styles. And yet, in the between-subjects design where individuals were exposed to only one explanation style repeated across multiple cases, these explanation effects largely disappeared.

One possible explanation is that when people repeatedly see the same explanation style, they become habituated and pay more attention to the features of the cases rather than any specific features of the explanation provided. When exposed to multiple explanation styles regarding the same case, the differences may become more salient. While one-off interactions with individual systems seem most likely, both scenarios---comparison of multiple styles, and exposure to a single style---may occur in different real world contexts. For instance, people may realistically face multiple explanations for similar algorithmic decisions when applying for a loan, mortgage or credit. Various explanation tools already exist within the credit scoring industry, and a savvy consumer might well test them all out when preparing for a significant application.

The relative consistency in justice-related judgements between different explanation styles might also be explained by strongly-held intuitions about the variables used in a decision. While different explanation styles represented the input variables in different ways, the case descriptions all featured the same set of inputs. If people have strong intuitions about the fairness of using particular features in a model (regardless of their weights), as has been observed in a previous study by Grgic et al~\cite{grgic2016case}, this may account for such consistency. Another important factor may be the extent to which experimental design allows interpersonal comparison. Research on justice judgements in non-automated contexts has found that individuals pay more attention to other people's outcomes than they do to information about the procedure behind the decision that affects them~\cite{van1997judge, xia2004price}.

Finally, the observed correlations between different justice constructs fall largely in line with those observed within the psychology of justice literature amassed since the 1970s. Procedural and distributive dimensions had a relatively high correlation coefficient of 0.69; for comparison, a 2001 meta-analysis of 45 studies in a variety of organisational and institutional settings reported a correlation coefficient of 0.48 for the same constructs~\cite{colquitt2001justice}.

\section{Limitations and Future Work}
While the combination of different methods employed in these three studies goes some way to addressing the problems associated with any single approach, there are a number of limitations which are important to mention.

\subsection{Threats to validity}
First, the samples were not representative of the general population. Participants for the in-person lab study were all drawn from an affluent and academically dominated city, and from the UK where experiences of certain decision contexts (e.g. financial loans) likely differ to other countries (e.g. those with greater or lesser dependence on private debt). Participants in the online studies were not gender balanced, with almost double the number of females to males. Second, the scenarios considered were hypothetical, not affecting the participants directly, and therefore lacked the first-person consequences and significance of a real world decision, as well as the possibility of alternative courses of action such as requesting a human review or appealing the decision. The explanations were also not drawn from real machine learning model outputs due to our desire to understand explanations in the context of rejections and not to introduce confounding elements---but in practice, this might introduce additional complicated factors. Finally, comparisons were made between four different explanation styles, but further confirmatory work may be needed to address the specific reasons for differences between explanation styles. In particular, the explanation styles proposed in the interpretable machine learning literature vary in the information they contain. Despite our efforts to standardise them, differences in the length and format of the text required to impart an explanation between explanation conditions may have been a confounding factor.

\subsection{Future work}
A number of opportunities for future work arise from this discussion and limitations. One would be to consider the application of other possible measures and methods from the psychology of justice research, such as interactional justice. Indeed, some participants' remarks (e.g. about the computer being 'rude') suggest that there may be useful notions of interactional justice even when humans are not directly involved in communicating the decision. Another would be to examine how to design systems designed to make ML outputs interpretable in different ways to multiple end-users for different ends; not only the data scientist attempting to detect and mitigate discriminatory effects (e.g.~\cite{berendt2014better}), but also other stakeholders such as third party auditors. This is part of a wider need to consider how to re-engineer better user experiences for multiple stakeholders interacting with machine learning systems in general~\cite{dove2017ux}.

\section{Conclusion}

Algorithmic decisions are likely to become increasingly relied on for a range of decisions with potentially important repercussions for those affected. Understanding how people assess the fairness of such decisions, and how explanations might help, is therefore of increasing significance. Despite repeated calls for more transparency over how such decisions are made, there is still much to learn about what people want and need to know about algorithms in order to hold them accountable to justice. As lawmakers legislate for mandatory provision of information to decision-subjects, human-computer interaction research has much to offer in how such information should be extracted, presented and delivered.

This paper suggests that people do consider justice-related aspects of algorithmic decision-making systems, much as they do for manual decision-making processes. However, depending on how and when they are deployed, explanations may or may not help individuals to evaluate the fairness of such decisions. The algorithmic nature of these systems results in an array of novel considerations which are not captured by traditional research on perceptions of justice. Conversely, creators of ML explanation systems have not typically designed them with the information needs of those individuals facing significant personal consequences of model outputs in mind. It is our hope that such concerns will instigate renewed focus on this range of important use cases for algorithmic explanations, and more broadly for HCI research to support the pursuit of justice as algorithmic decision-making systems take hold in a wide array of high-stakes domains.

\section{Acknowledgments}
All authors were funded by the Engineering \& Physical Sciences Research
Council (EPSRC), under SOCIAM: The Theory and Practice of Social Machines EP/J017728/2 (Oxford), and EP/M507970/1 (UCL).
\bibliographystyle{SIGCHI-Reference-Format}
\balance
\bibliography{sample}

\end{document}